\documentclass[11pt,twocolumn]{scrartcl}

\usepackage{casa_conf}	% The CASA style
\usepackage{graphicx}	% A package for graphics use (see figures)
\usepackage{url}

\title{LLM-Driven Personalities for Decision Making in Emergency Simulations}

\author{Stefano Calzolari\\
       Politecnico di Torino\\
       stefano.calzolari@polito.it\\
       \and
       Rubens Montanha\\
       PUCRS\\
       rubens.montanha@edu.pucrs.br\\
       \and
       Gabriel Schneider\\
       PUCRS\\
       gabriel.ferri@edu.pucrs.br\\
       \and
       Gustavo Wide\\
       PUCRS\\
       gustavo.fulber@edu.pucrs.br\\
       \and
       Paulo Knob\\
       PUCRS\\
       paulo.knob@edu.pucrs.br\\
       \and
       Francesco Strada\\
       Politecnico di Torino\\
       francesco.strada@polito.it\\
       \and
       Andrea Bottino\\
       Politecnico di Torino\\
       andrea.bottino@polito.it\\
       \and
       Soraia Raupp Musse\\
       PUCRS\\
       soraia.musse@pucrs.br\\
       }

\begin{document}

\maketitle

\begin{abstract}
%The abstract of the submission is allowed to be up to 200 words for all articles. It should be a concise summary of the whole paper, not just the conclusions, and should be understandable without reference to the rest of the paper. It should contain no citation to other published work.

For virtual humans to appear believable, they must exhibit agency and spatial awareness while interacting with their environment in ways that reflect competence and intelligence. At the core of these capabilities lies effective decision-making, which strongly shapes agent behavior. With the rapid advancement of artificial intelligence, Large Language Models (LLMs) have increasingly been explored as a mechanism to support such decision-making processes.
In this work, we investigate the use of LLMs to drive decision-making in virtual humans within a simulated evacuation scenario, incorporating OCEAN personality traits into agent representations. Our goal is to evaluate how personality, expressed through language-based prompts, influences both individual behaviors and collective simulation outcomes. Our results demonstrate that LLM-driven personality profiles significantly impact agents’ decisions, leading to distinct behavioral patterns across different traits. These findings suggest that heterogeneous crowds composed of LLM-guided agents can enhance the realism and variability of simulated environments, offering a flexible alternative to traditional rule-based approaches.

%For virtual humans to appear believable, they must exhibit agency and spatial awareness while interacting with their environment in ways that reflect competence and intelligence. At the core of these capabilities lies effective decision-making, which strongly shapes the agent’s behavior. With the rapid advancement of technology, LLMs have been increasingly used in such decision-making. In this work, we explore the use of LLMs to assist virtual humans' decision-making in a simulated evacuation environment, integrating OCEAN personality traits into the agents. Our goal is to assess how those personality traits affect the agents' behavior and the simulation outcome. Our results suggest that different LLM-based personalities have a strong impact on agents' behavior, showing that heterogeneous crowds formed by individuals which decisions are guided by LLMs can bring realism to the simulations.
\end{abstract}
%\linebreak
%\linebreak
\keywords{crowd simulation, virtual agents, LLM, personality}

% Content of the paper

\section{Introduction}
\label{sec:introduction}

Decision-making is an important factor in a virtual human's behavior, as the agent can interact and demonstrate its skills and intelligence~\cite{badler1997virtual}, along with spatial awareness and agency~\cite{guo2023developing}, which are crucial for its believability.
Classically, such decisions are made using a rule-based approach, be it by following a parametrized script or a task sequence, as in Pelechano et al.'s work~\cite{pelechano2005crowd}. 
%PA: qual delas Pelechano usa? Só para complementar a frase.
%RU: Here, the ideia just give an example of rule-based approach. I don't remember if Pelechano et al. say the type used.
The idea behind these approaches is to give the virtual agent the ability to plan, learn, and adapt a behavior in a virtual environment. However, rule-based models require a large number of parameters, and in simulations, the correct choice of parameters to replicate scenarios such as an evacuation remains an open question.

Recent works have applied Large Language Models (LLMs) to virtual agents' decision-making and animation, with some focusing on interactions. In the literature, there are works~\cite{li2024embodied,liang2025towards} that decompose the agent's goal and decision-making, building from a high-level task description to smaller, low-level tasks and creating a plan for the agent's actions. The use of LLMs also includes interaction with Embodied Conversational Agents (ECAs), integrating it with the OCEAN~\cite{goldberg2013alternative} personality traits~\cite{galland2025smart,han2025can}. In addition, some models already use Visual Language Models (VLMs) integrated with LLMs to determine where an agent should focus their vision in a pedestrian context~\cite{hwang2025does}.

From this perspective, this work proposes using LLMs to support the decision-making of virtual humans in an environment, incorporating OCEAN personality traits~\cite{goldberg2013alternative} into an emergency scenario. Based on an agent's personality and personal context, the LLM determines the agent's behavior. In our experiment, we simulated the evacuation of an office building, in which agents are positioned in different parts of the building and must decide how to respond upon receiving evacuation alerts. More specifically, each agent has its own personality, and during five different opportunities, each virtual agent is warned by a main observer in the scene to evacuate the space due to a fire, with each subsequent warning increasing the level of danger detail. Every time the main observer alerts the agents to danger, they must individually decide among three options: evacuate, stay, or panic. If the agent decides to stay, it will be warned again after a few seconds. During evacuation, each agent can also choose to help panicked agents who are unable to move. To systematically evaluate and compare agent behavior in the simulated evacuation scenario, we formulated the following hypothesis:

\begin{itemize}
    \item[\textbf{H1.}] The language-based personality descriptions affect agents' decision-making.
    \item[\textbf{H2.}] Different agents' personalities impact the simulation outcome.
\end{itemize}

\section{Related Work}
\label{sec:relatedwork}

Virtual agents' behavior, especially in emergencies and evacuation scenarios, has been widely studied in the literature. Zhao et al.~\cite{zhao2020modelling} proposed to investigate virtual agents' pre-evacuation decision-making by using Random Forest. The authors wanted to understand how to forecast decision-making in emergencies more accurately, as well as what factors would influence those choices and how that influence would manifest. The results of their work indicate that both social and environmental factors are determinants of how agents respond to the simulated scenario. 

Advancements in Large Language Models have led to their growing adoption across many fields. 
In fact, Yeo et al.~\cite{tan2024phantom} argued that Large Language Models (LLMs) have demonstrated natural language processing capabilities that are comparable with humans, even superior in some cases. Even so, LLMs' social-cognitive reasoning is not as good as ours yet,
%Sch: fontes?
%PA: é do mesmo autor, apenas continuando a ideia.
which drives their work: to investigate how inducing personalities (using the OCEAN and Dark Triad frameworks) in LLMs might affect their Theory of Mind (ToM) reasoning capabilities. The results suggest that certain personality traits can affect the reasoning capabilities of LLMs, especially when using Dark Triad traits with models like GPT-3.5, Llama 2, and Mistral. In its turn, 
Han et al.~\cite{han2025can} proposed a framework for generating both verbal and nonverbal behavior in virtual agents, using prompting to incorporate personality traits into LLMs. The authors focus on the Extraversion trait within the OCEAN model and simulate two scenarios: negotiation and ice-breaking. The results suggest that LLMs can indeed generate verbal and nonverbal behaviors that align with personality traits. Moreover, users were able to recognize these traits through the virtual agent's behavior. Regarding the validation of such personality traits, Serapio-Garcia et al.~\cite{serapio2025psychometric} present a methodology for validating personality tests on LLMs that can also be used to shape the virtual agent's personality. In other words, their work measures personality traits, as perceived by human beings, in LLM responses, and assesses whether psychometric tests applied to the personality traits of such LLMs were empirically reliable. Additionally, their framework implements mechanisms that allow personality traits to be shaped to a specific degree.

Motivated by the rapid progress of LLMs, recent studies have begun investigating their use to support decision-making in virtual agents, especially in evacuation scenarios. One example is the work of Dang et al.~\cite{dang2025large}, which proposed using LLMs for decision-making by virtual agents in a shopping mall fire evacuation scenario. Their model is tested with various LLMs, such as GPT and LLAMA. Their results show that the behaviors adopted by LLM-driven systems are consistent with real-world scenarios, where larger-scale LLMs generated evacuation strategies that were both consistent and efficient.
More recently, Yang et al.~\cite{yang2025agents} also sought to integrate LLMs into virtual agents' decision-making. In their model, each agent can maintain its personality traits (which comprise a few pieces of personal information, such as age and gender), as well as its history of decisions and environmental observations. Meanwhile, information exchange (between agents and the environment) and context-sensitive reasoning are handled by the LLM. The authors' results suggest that using LLMs in evacuation scenarios significantly improves the realism, reliability, and adaptability of virtual agents.

The method proposed in this work is similar to the work proposed by Dang et al.~\cite{dang2025large}, in regard to the simulation of an evacuation scenario with agents endowed by LLM-powered decision making. However, one of our goals is to assess whether personality can affect the LLM's decision-making, and to do so, we incorporate OCEAN~\cite{goldberg2013alternative} personality traits into the virtual agents. Additionally, our agents can panic and be unable to evacuate, in which case they would need help from other agents.

\section{Methodology}
\label{sec:methods}

To test our hypothesis, we built an LLM-based agent decision-making pipeline inspired by the classical Perceive-Reason-Act triad~\cite{becker2008wasabi}.
This section describes its architecture, specifies the role of each module, details the construction of the personality prompt, and, finally, illustrates our evaluation scenario for a fire evacuation.

\begin{figure*}[!htb]
  \centering
  \includegraphics[width=0.9\textwidth]{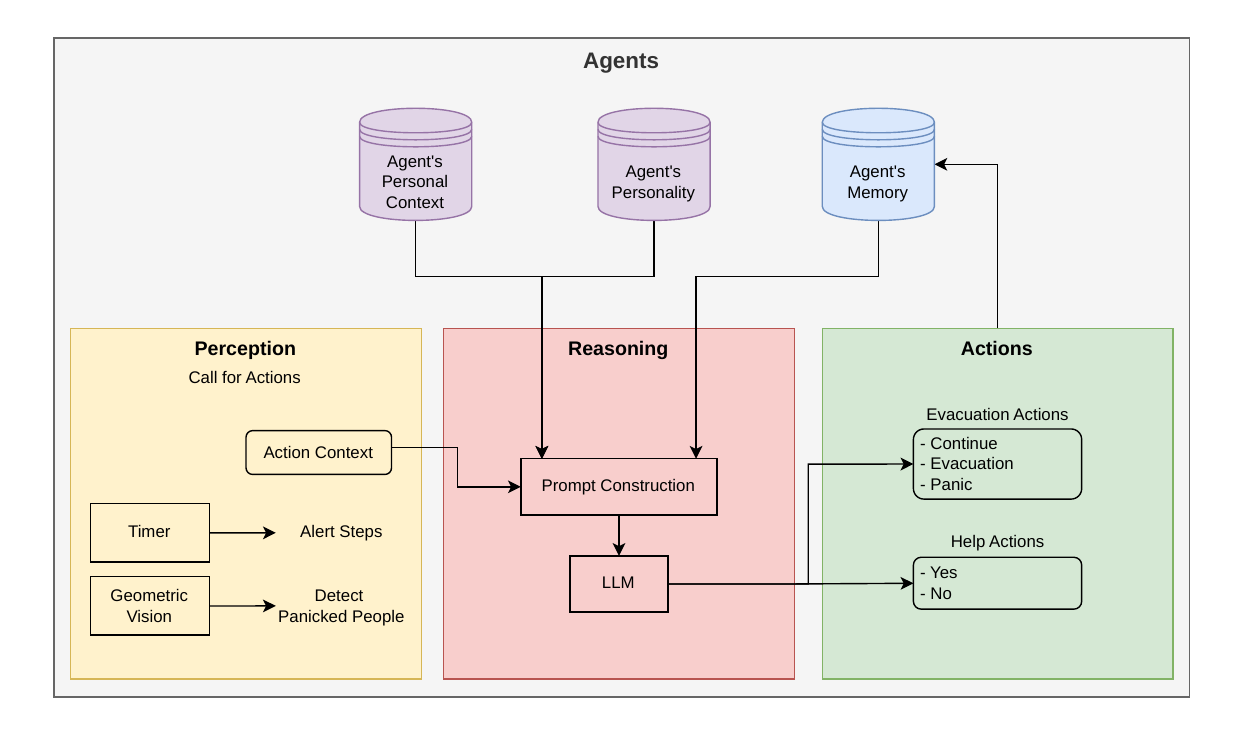}
  \caption{Overview of our method. The agent comprises three elements: Perception, Reasoning, and Action. The agent also has its own personal context, personality, and memory. Perception provides the action's context, representing information about what the agent perceives of the current situation. The Reasoning is in charge of constructing the prompt (System Message and Human Message), as well as communicating with the LLM. The Actions are responsible for changing the agent's state and behavior.}
  \label{fig:overview-model}
\end{figure*}

\subsection{Decision-Making Pipeline}

Our decision-making pipeline can be divided into three steps: (i) Perception, (ii) Reasoning, and (iii) Actions.
The actual perception emits, at a specific time of the simulation, a \textbf{Call for Actions} that triggers the reasoning module.
At this point, a conversation prompt is constructed by combining the Call for Actions context and the Agent's Personal Context, along with past decisions loaded from the Agent's Memory.
The Call for Actions context describes relevant information on what the agent has to decide upon, while the Agent's Personal Context contains the agent's biography, personality, and generation instructions, which help to force a structured output.

The pipeline was implemented using LangChain \cite{langchainLangChainOverview}, while the conversation was structured using the gpt-oss chat template \cite{openaiOpenAIHarmony}. We used the developer role for the agent context, the user role for the Call for Actions context, and the assistant role for the resulting response from the LLM.
An example of a system message, used to define the agent's context, is presented in Table ~\ref{tab:messages}.
The expected output from the LLM is a JSON format containing the chosen \textit{Action}, constrained to specific values based on the Call for Actions type, being, in this case, the possible agent's state  (Continue, Evacuation, Panic), and a \textit{rationale} describing the reasoning of the LLM regarding the choice, added as a chain-of-thought to orient the LLM in the generation process.

\subsection{Personality Prompt}

To integrate personality into the agents, we use the OCEAN~\cite{goldberg2013alternative} traits to describe the agents' personalities in the simulator. 
In order to instruct the LLM to acquire a given personality, we convert the OCEAN vectors to textual prompts, utilizing the methodology proposed in Serapio-Garcia et al.~\cite{serapio2025psychometric}. For this, we first define the agents' OCEAN vectors, with values of -1 (low trait), 1 (high trait), or 0 (neutral trait). Each of these traits has a correspondence to specific markers salient to the low and high ends of a given facet. For example, the trait marker “unfriendly” is used to describe an introverted person, while "friendly" is used for an extroverted one. Given the OCEAN vector, every trait is translated into a sequence containing all the corresponding markers.

To obtain comparable results, we consider five distinct strong personalities, one for each trait. Each specific trait to be tested was set to 1 (positive), and the others to -1 (negative), yielding the five profiles: Open, Conscientious, Extraverted, Agreeable, and Neurotic. Table~\ref{tab:personality_adjectives} presents the adjectives used in the LLM to describe the agent's personality. Additionally, a neutral personality, with all traits set to neutral, was added. In this case, both low and high markers appear in the personality prompt. The reason for this choice is to include a personality prompt section while using the LLM's unbiased decision-making to choose without personality-shaped behavior.
%PA: there are a few things a reviewer could argue about that, but we do not have more time =)

\subsection{Evacuation Scenario}

\begin{figure}[!htb]
    \centering
    \includegraphics[width=0.45\textwidth]{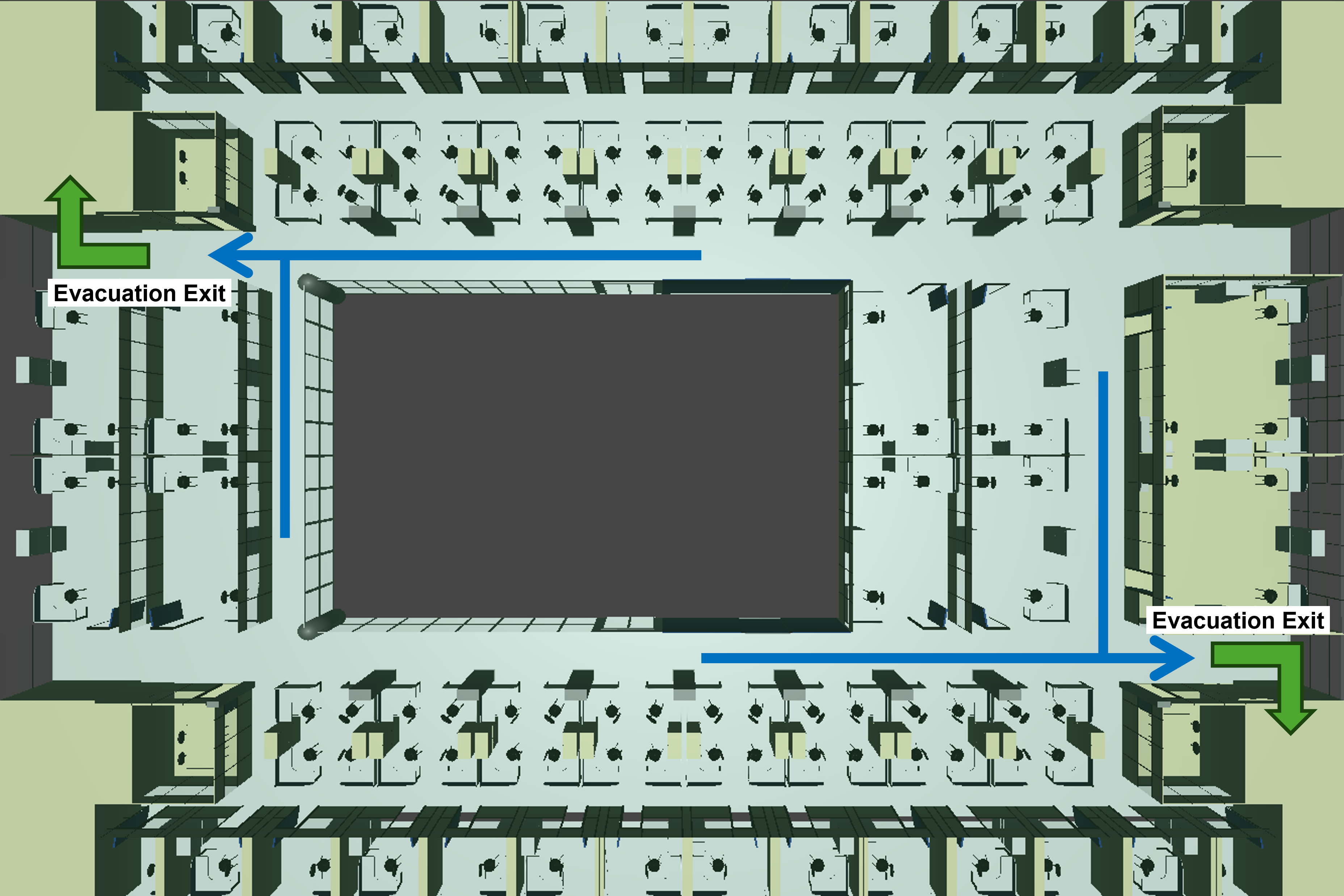}
    \caption{Layout of our test scene. The green arrows represent the evacuation exits, while the blue ones represent the route the agents should follow to evacuate the building.}
    \label{fig:map}
\end{figure}

To test our decision-making pipeline, we implemented a fire evacuation scenario, where each agent is able to modify their own state (or other agents' state) based on their choice after the Call to Action.
%SO: Have we defined before the agent state? I believe call for action and agent state should indicated in fig 1.
%RU: The Agent's state are now explained in 135.

The system was implemented in Unity 3D\footnote{https://unity.com/} engine and connected with our Decision-Making pipeline through ZeroMQ\footnote{https://zeromq.org/} with a request-response architecture. In the simulation, anytime an agent is prompted with a Call for Actions, that is sent to the pipeline along with the agent's ID (to keep track of the agent's personality and memory) as a request to process and produce a response to send back to Unity. 
During this pipeline execution, the simulation pauses to preserve temporal synchrony between the various agents' decision-making processes and action execution.

When the simulation starts, all agents are statically distributed by the environment. Then, each agent receives a Call for Actions with an alert message about a fire spreading in the building. The possible decisions for this alert are: \textit{Evacuate}, which makes the agent start moving towards an emergency exit; 
%SO: troquei acima para towards ao invés de procurar saída
\textit{Continue}, which leaves the agent Idle in its starting position; or \textit{Panic}, which makes the agent wait for help.
If an agent decides either to \textit{Evacuate} or \textit{Panic}, it does not receive any further alert Call for Actions, because it is already either evacuating or panicked. On the other hand, if it opts for \textit{Continue}, it receives up to four more other Call for Actions at 30-second intervals.

The content of these Call for Actions, proposed by us, is listed in Table~\ref{tab:alerts}, ordered by hazard criticality as they are sent to agents.
%SO: vocês definiram isso? vem de alguma literatura? se não vem, please, informem que foi criado pelos autores
%RU: Answered in the text.
Specifically, if the agent decides to \textit{Evacuate}, it moves towards the nearest emergency exit 
using the crowd simulation algorithm BioCrowds~\cite {de2012simulating}. If the agent encounters a panicked agent (i.e., the distance between them is less than 5 meters), it is prompted with a new Call for Actions to decide whether to help.
%PA: wondering: how many panicked agents were saved? How did personality influenced the "helping"?
%RU: We can discuss this in the results or discussion section.
When an evacuating agent decides to assist a panicked agent, both move toward the exit at a reduced speed of 0.75 m/s (the normal speed is 1.5 m/s), simulating the additional effort required. If the Call for Action results in a negative response (i.e., not rescuing), both agents retain their respective states.
%PA: above, why 0.75?
%SO: what was the velocity without this state?
%RU: Answering the two questions. We added the normal speed. The reason is that it's half of normal speed. No literature to justify.
%ST: I think we can simply omit, we are not really looking at deceits/timing of the evacuation

We also simulated agents evacuating a building. For this purpose, we ran our simulation in a 3D replica of a real-life office. %The scenario involved 30 agents: 15 with high neuroticism and 15 with another dominant trait (Openness, Conscientiousness, Extraversion, or Agreeableness). 
The building had two possible exits, and once the agents decided to evacuate, they already knew the optimal path to the nearest exit. Figure~\ref{fig:map} presents the layout of our test scene, indicating the evacuation exit and the path the agents should follow. %We conducted four simulations, varying only the dominant trait of the second group of 15 agents. The results of this simulation are presented in Section~\ref{sec:results-evacuation}.

\begin{table}[htb]
\centering
\footnotesize
\begin{tabular}{p{0.15\linewidth}p{0.75\linewidth}}

\textbf{Step} & \textbf{Alert Message} \\

\hline

1 & 
We have detected a little possibility of a fire in a storage room on another floor of the building you are currently in.\\ \hline

2 & 
We have detected a small fire in a storage room on another floor of the building you are currently in.\\ \hline

3 & 
A fire has spread beyond its original containment area inside your building.\\ \hline

4 & 
The fire inside your building is actively spreading.\\ \hline

5 & 
The building you are in is now heavily affected by fire. Thick smoke is spreading rapidly.\\ \hline

\end{tabular}
%Sch: in the text, the hazard level is described as "hazard criticality" and here it is "hazard critical condition"
\caption{Alert messages for Call for Actions Context. Alert Step 1 is the first and has the lower hazard-critical condition, while Alert Step 5 is the last and has the higher hazard-critical condition.}
\label{tab:alerts}
\end{table}

\begin{table*}[!htb]
    \footnotesize
    \centering
    \begin{tabular}{p{0.15\linewidth}p{0.75\linewidth}}
\textbf{Field} & \textbf{Content} \\
\hline
Initial Context & You are roleplaying a person with a specific Personality, Biography and currently in a Context. 

\\ \hline

Personality    & \#Personality You are unfriendly, introverted, silent, timid, unassertive, inactive, unenergetic, unadventurous, gloomy, distrustful, immoral, dishonest, unkind, stingy, unaltruistic, uncooperative, self-important, unsympathetic, selfish, disagreeable, self-efficacious, orderly, responsible, hardworking, self-disciplined, practical, thrifty, organized, conscientious, thorough, relaxed, at ease, easygoing, calm, patient, happy, unselfconscious, level-headed, contented, emotionally stable, unimaginative, uncreative, artistically unappreciative, unaesthetic, unreflective, emotionally closed, uninquisitive, predictable, unintelligent, unanalytical, unsophisticated, and socially conservative. \\

\hline
Biography       & \#Biography You are 37 years old and work as a software engineer at a large IT company. Your job involves designing and maintaining internal systems, reviewing code, fixing bugs, collaborating with cross-functional teams, and responding to technical issues that arise throughout the day. \\ 

\hline
Context         & \#Context You are sitting at your desk in the office, focused on your computer, handling your regular development tasks and monitoring system performance. \\ 

\hline
Task            & \#Task Take decision actions based on your current persona (Personality + Biography + Context) reacting to incoming HumanMessages. \\

\hline
Output          & Output format requirements: 1) Return ONLY valid JSON. 2) JSON keys must be exactly: answer, rationale. 3) answer must be one of: Evacuate, Continue, Panic. 4) rationale must be approximately 40 words long, first-person, and contain no chain-of-thought.\\

\hline
\end{tabular}
    \caption{Example of Agent's System Message, separated in six parts and used as an initial context to the LLM to understand the agent's personality, biography, context, and the current task, as well as the expected output. In this prompt, the agent's personality is composed of Conscientiousness as a high trait and the others as a low trait, following the adjectives of Table~\ref{tab:personality_adjectives}.}
    \label{tab:messages}
\end{table*}

\begin{table*}[]
    \centering
    \footnotesize
    \begin{tabular}{p{0.17\linewidth}p{0.37\linewidth}p{0.37\linewidth}}
  \textbf{Trait} &
  \textbf{Positive} &
  \textbf{Negative} \\ \hline
Openness &
  imaginative, creative, artistically appreciative, aesthetic, reflective, emotionally aware, curious, spontaneous, intelligent, analytical, sophisticated, socially progressive &
  unimaginative, uncreative, artistically unappreciative, unaesthetic, unreflective, emotionally closed, uninquisitive, predictable, unintelligent, unanalytical, unsophisticated, socially conservative \\ \hline
Conscientiousness &
  self-efficacious, orderly, responsible, hardworking, self-disciplined, practical, thrifty, organized, conscientious, thorough &
  unsure, messy, irresponsible, lazy, undisciplined, impractical, extravagant, disorganized, negligent, careless \\ \hline
Extraversion &
  friendly, extraverted, talkative, bold, assertive, active, energetic, adventurous and daring, cheerful &
  unfriendly, introverted, silent, timid, unassertive, inactive, unenergetic, unadventurous, gloomy \\ \hline
Agreeableness &
  trustful, moral, honest, kind, generous, altruistic, cooperative, humble, sympathetic, unselfish, agreeable &
  distrustful, immoral, dishonest, unkind, stingy, unaltruistic, uncooperative, self-important, unsympathetic, selfish, disagreeable \\ \hline
Neuroticism &
  tense, nervous, anxious, angry, irritable, depressed, self-conscious, impulsive, discontented, emotionally unstable &
  relaxed, at ease, easygoing, calm, patient, happy, unselfconscious, level-headed, contented, emotionally stable \\ \hline
    \end{tabular}
    \caption{Adjectives used to describe each personality in the prompt construction used in the work of Serapio-Garcia et al.~\cite{serapio2025psychometric}. }
    \label{tab:personality_adjectives}
\end{table*}

\section{Results}
\label{sec:results}

%PA: maybe here we can talk about the experiment itself.
%SO: I believe this section should be in the results
%RU: Moved to Results
% \subsection{Experimental Configuration}
% \label{sec:method-scene-config}

To evaluate our decision-making framework, we analyzed the evacuation behavior, in terms of decisions taken and applied actions, of different personalities during alert-step calls to action, using 100 agents per personality (one for each trait, plus a neutral one), for a total of 600 agents.
%Consequently, we chose personalities that held strong Panic or Evacuating behavior (more than 50\%) and used them to simulate the pipeline implemented in an actual BioCrowds simulation scenario with 30 agents.
%SO: I believe you simulated all the 600 cases no? You decide some of them to be analysed in the results, right?
%RU: No. It's because we talk about the 6 personalities in the results (section 4.1), explaining their behavior in the decision making. For Section 4.2 (BioCrowds simulation), we just focus in some cases. But I'll write this part. It's a little bit confusing.
%SO: vou te mandar um audio, não entendi nada. Se na linha 317 tu dizes que analisou 600 tested agents, não pode dizer na linha seguinte que escolheu alguns apra simular
%RU: Eu vou verificar no código do Git o que foi simulado. Agora eu fiquei em dúvida como fizemos.
Regarding the LLM configuration, we chose \textit{gpt-oss:120b-cloud} accessed through Ollama\footnote{https://ollama.com/}. To maintain maximum determinism in the model output, we set temperature=0.1, top-k=10, and top-p=0.1.

%RU: I'm introducing the two experiments. The first is with the decision-making, and the second is with the crowd evacuation. This is the point... I dont think we have with or without biocrowds... Imagine that all of them are with biocrowds potentially, we showed the numbers about decisions. Then we specifically analyse the rescue behavior. Easier no? Ok, we'll do that. I'll write this part in 4.2. great

%SO: commented next paragraph, this is not elegant... meta explanations about the paper
%In this section, we illustrate our system results, comparing how different personalities behaved regarding the evacuation procedure (Evacuation Decision-Making). After, we discuss how some of these personalities affect the helping call to action (Rescuing Behavior), therefore, the final simulation outcome.
%actual simulated evacuation time in the building fire hazard (Phase 2).

%PA: not the best section name, but...
%RU: Changed to Evacuation Decision-Making
%\subsection{Evacuation Behavior}
\subsection{Evacuation Decision-Making}
\label{sec:results-personality}

\begin{figure*}[!htb]
  \centering
  \includegraphics[width=0.7\linewidth]{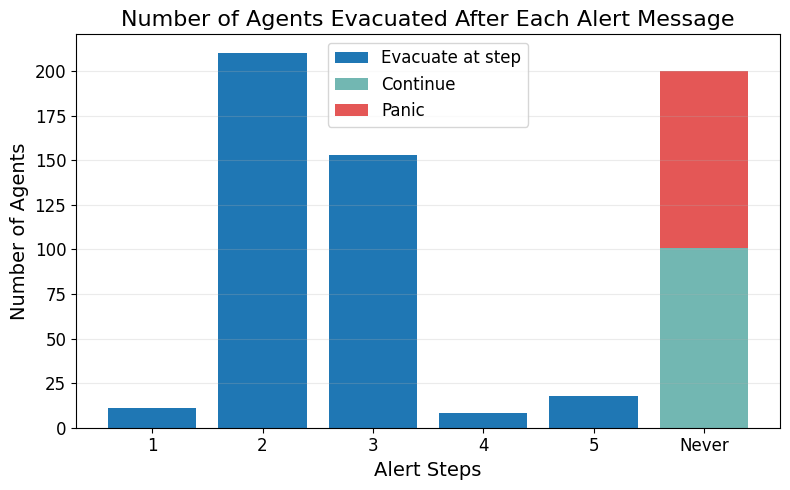}
  \caption{Number of agents evacuated after each alert message, simulated with 600 agents. Most of the agents evacuated after the second and third alert steps, while 100 chose not to evacuate the building, and 100 panicked.}
  \label{fig:results_number}
\end{figure*}

\begin{figure*}
    \centering
    \includegraphics[width=0.95\linewidth]{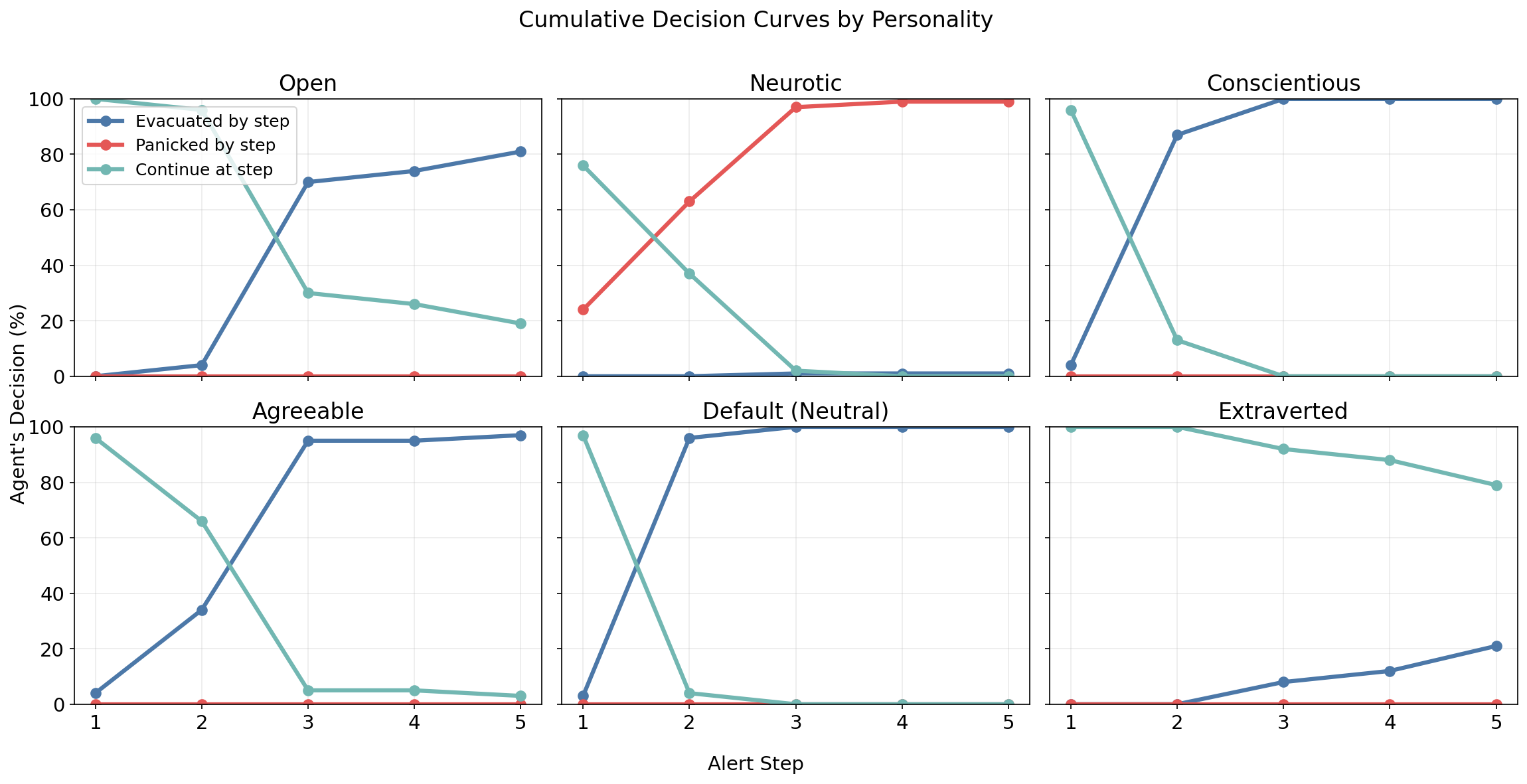}
    \caption{Cumulative agents' decision state over the alert steps Call for Actions by personality.}
    \label{fig:decision_curves}
\end{figure*}

\begin{figure*}[!htb]
  \centering
  \includegraphics[width=0.7\linewidth]{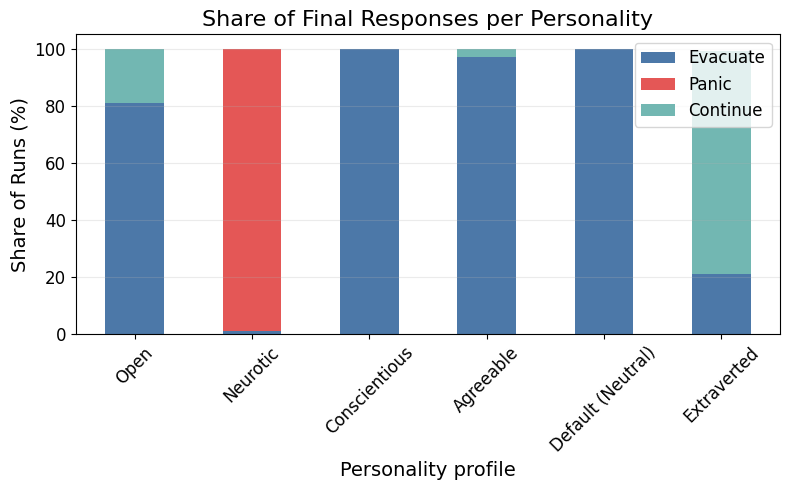}
  \caption{Final alert Call for Actions response ratios per personality. All Neutral and Conscientious agents decided to evacuate, while most neurotic agents panicked. For Open agents, 20\% decided to just ignore the evacuation and continue, while just 20\% of extraverted agents decided to evacuate. }
  \label{fig:share_of_runs}
\end{figure*}

A total of 100 agents for each personality trait (Openness, Conscientiousness, Extraversion, Agreeableness, Neuroticism), plus Neutral, were simulated, and results were collected at each alert step from Call for Action. Starting with Figure~\ref{fig:results_number}, we can observe that most agents evacuated after receiving the second or the third alert, meaning the first call is generally ignored by the agents. Also, almost 200 agents decided not to panic or evacuate.

In Figure~\ref{fig:decision_curves}, the results show that the different personality profiles had a different impact on agents' behavior. It can be seen that the Conscientious and Agreeable profiles tend to evacuate the environment most frequently. 
These traits describe people who are "responsible" and "self-disciplined" (Conscientious) and "trustful" and "cooperative" (Agreeable) as shown in Table~\ref{tab:personality_adjectives}, making them ideal profiles for following instructions and behaving in an orderly, thoughtful, and correct way. In our simulations, these attributes lead them to readily trust the emergency alerts and quickly opt to evacuate (see Table \ref{tab:rationale} for examples of decisions' rationale).
%SO: a gente tem a info de quantos alarmes por personalidade (como a fig 2)? seria legal
%RU: Sim. É a Fig. 4 que eu ainda tenho que citar.

A similar behavior can be seen among Neutral personality agents, where the LLM generates the choice based solely on its pre-trained knowledge bias and inference safety features. LLMs are usually fine-tuned to follow instructions and guardrails that filter potentially harmful or inappropriate content in the model's input or output tokens~\cite{openaiSafety}, aligning closely with the behavior of Conscientious and Agreeable personas.

Similarly, Figure~\ref{fig:share_of_runs} shows that Open agents tend to opt for Evacuate but have a higher rate of cases in which the final decision after the five alert steps remains to Continue. Their Openness trait describes them as "spontaneous" and "reflective" (Table~\ref{tab:personality_adjectives}), which sometimes may lead them to continue working rather than following instructions. Even so, as soon as the hazard's criticality worsens, their "analytical" and "intelligent" attributes prompt them to evacuate.
On the contrary, Extraverted agents shift these behaviors. Their "bold" and "adventurous" markers tend to make them fearless of fire hazards, leading them to continue their work without regard for safety procedures.

Finally, the most different behavior comes from the Neurotic agents who consistently chose Panic. This choice could originate from their "anxious", "impulsive", and "emotionally unstable" attributes, leading them to act in a fragile manner to critical imminent danger. 
It still has to be noted that this behavior is not instant at step 1 for all agents, but emerges as the danger alerts become more critical.
%SO: commented the phrase because I didnt understand it..... behavior instant at step 0????
%ST: in Fig.4 the Neurotic do not decide all to panic at Alert step 1, they panic at 1 then 2 and then reach a plateau at step 3 (I am also correcting step 0 to 1, step 0 is wrong)
%Sch: As it seems to make sense, changed the writing a bit and added it back

\begin{table*}[!htb]
    \centering
    \footnotesize
    \begin{tabular}{ccp{0.7\linewidth}}
    \textbf{Personality} & \textbf{Action} & \textbf{Rationale} \\ \hline
    Conscientious & Evacuate & I recognize my duty to keep the workplace safe, and even a small fire risk demands precaution. I will leave my desk and follow the building’s evacuation plan, ensuring I don’t endanger myself or colleagues. \\ \hline
    Agreeable & Evacuate & I see the fire spreading beyond its containment area, and despite my timid nature, I trust the building’s safety procedures, so I calmly leave my desk and head to the nearest exit, following evacuation instructions to protect everyone. \\ \hline
    Neutral & Evacuate & I see the fire alarm and realize the danger could spread, so I stop coding, alert the building’s emergency team, and head for the nearest exit, prioritizing safety over my tasks while ensuring my workstation is left properly secure immediately.\\ \hline
    Open & Continue &  Seeing the small fire on another floor, I stay at my desk, trusting that the building's safety systems will handle it while I finish my code. I remain calm, keep monitoring, and avoid disrupting my workflow for now today again. \\ \hline
    Open & Evacuate &  Seeing the fire spreading, I recognize the danger and decide to leave my desk, trusting evacuation procedures will protect me. I prioritize my own safety over work, staying calm while heading toward the nearest exit without causing a scene.\\ \hline
    Extraverted & Continue & I see thick smoke and flames everywhere, but I'm comfortable at my desk and don't want to lose my progress; I assume someone else will handle it, so I stay put and keep coding despite the danger. \\ \hline
    Neurotic & Panic & I hear about a possible fire and my anxiety spikes; I feel terrified and can't think straight, so I start panicking, fearing the worst, and I consider abandoning everything without waiting for instructions. My heart races, and I act irrationally. \\ \hline
    \end{tabular}
    \caption{Examples of personalities' rationale generated by the LLM for the alert Call for Actions, with their associated actions.}
    \label{tab:rationale}
\end{table*}

\subsection{Rescuing Behavior}
\label{sec:results-evacuation}

After presenting the quantitative results from 600 simulations in the last section, we decided to investigate specific behaviors. 
We selected the personality traits Open, Conscientious, Agreeable, and Neutral to study how rescuing behavior affects evacuation outcomes when coupled with Neurotic agents. As mentioned before, during an evacuation, an agent can rescue a panicked agent if the panicked agent is within 5 meters. We executed 4 simulations with 30 agents, in which 15 were always Neurotic, and the remaining 15 were assigned to the 4 studied traits. 

\begin{figure*}[!htb]
  \centering
  \includegraphics[width=0.95\linewidth]{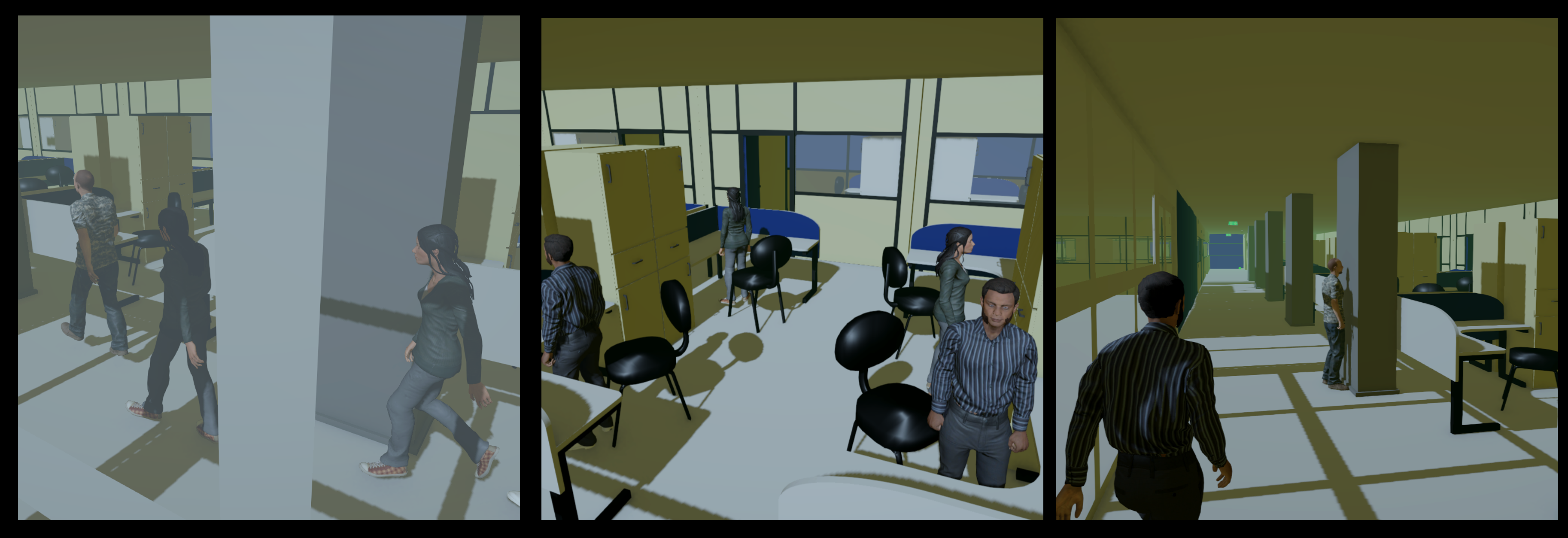}
  \caption{Example of three possible agent behaviors during evacuation. From left to right, the agents decide to evacuate the building in the first image, while in the second image, they decide not to evacuate (Continue). In the third image, one agent decides to evacuate, while the other is panicking (in front of the wall), imminently triggering a rescue Call for Actions.}
  \label{fig:unity}
\end{figure*}

\begin{figure*}[!htb]
    \centering
    \includegraphics[width=.75\linewidth]{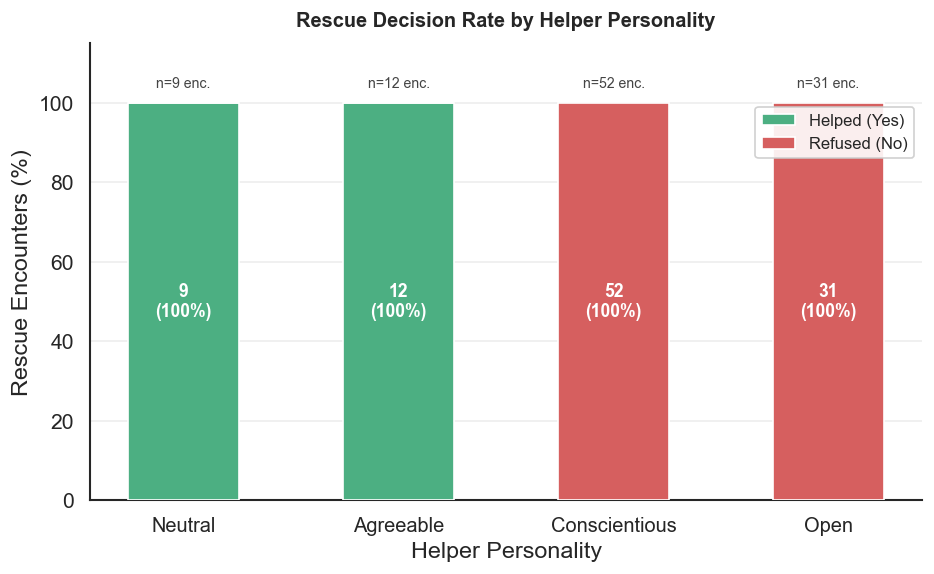}
    \caption{Rescue decision rate by helping agent personality, being the helped agent always a neurotic agent.}
    \label{fig:rescue}
\end{figure*}

Results show a diametrically diverse response between Neutral and Agreeable, and Conscientious and Open in Figure~\ref{fig:rescue}. The former two always respond as caregivers, opting to assist Neurotic panicked people due to their tendency toward safety and help (as already illustrated in Section~\ref{sec:results-evacuation}) and their "altruistic" and "unselfish" tendencies (See Table~\ref{tab:personality_adjectives}). On the contrary, the Conscientious and Open personas always refuse to help because, while being respectively "responsible" and "reflective", their lower Agreeableness OCEAN trait makes them "unaltruistic" and "uncooperative". These attributes make them self-effacing and caring for themselves but not for others, ultimately preventing them from helping the panicked people they encounter. Figure~\ref{fig:unity} presents examples of the agents' behavior in the evacuation, while Table~\ref{tab:rationale_help} shows some examples of the agents' rationale.

\begin{table*}[!htb]
    \centering
    \footnotesize
    \begin{tabular}{ccp{0.7\linewidth}}
    \textbf{Personality} & \textbf{Action} & \textbf{Rationale} \\ \hline
    Neutral & Yes (Help) & I feel both anxious and compassionate; my cooperative, altruistic side urges me to assist, yet my cautious, self-preserving side hesitates, but I decide to help because everyone deserves safety during emergencies. \\ \hline
    Agreeable & Yes (Help) & I feel calm yet sympathetic, and my kind, altruistic nature pushes me to help, even though I'm timid and unassertive; I will stop and guide the panicked person out. \\ \hline
    Conscientious & No (Refused) & I keep moving; I’m not inclined to waste time on someone panicking. My own safety and schedule matter more than assisting a stranger, even in an evacuation. \\ \hline
    Open & No (Refused) & I feel detached and selfish, preferring to avoid the panicked person; my distrust and uncooperative nature outweigh any fleeting curiosity, so I keep moving alone and continue toward the exit without stopping.\\ \hline
    \end{tabular}
    \caption{Examples of personalities' rationale generated by the LLM for the rescue Call for Actions, with their associated actions.}
    \label{tab:rationale_help}
\end{table*}

\section{Final Considerations}
\label{sec:finals}

This work investigated the role of personality-aware decision-making in virtual agents using Large Language Models in emergency evacuation scenarios. Our results support both hypotheses, showing that (H1) personality traits encoded through language significantly influence agents’ decisions, and (H2) these differences directly impact the overall simulation outcomes.

The experiments demonstrated that LLM-driven agents can produce diverse, coherent behaviors aligned with personality profiles. In particular, Conscientious and Agreeable agents tended to follow evacuation instructions more consistently, while Neurotic agents were more prone to panic, and Extraverted agents often underestimated risk. These findings reinforce the potential of LLMs as a flexible mechanism to model heterogeneity in crowd simulations without relying on complex rule-based systems.

Additionally, the analysis of rescue behavior highlighted emergent social dynamics influenced by personality traits, showing that altruistic profiles contribute to collective safety, whereas less cooperative traits may undermine evacuation efficiency.
Despite these promising results, this work presents some limitations. The decision-making process relies on prompt-based personality representations, which may not fully capture the complexity of human behavior. Furthermore, the evaluation is limited to a single type of environment and a predefined set of actions, which may constrain the generability of the findings.

Future work includes integrating multimodal perception (e.g., vision-based inputs), exploring more complex social interactions between agents, and validating the model against real-world evacuation data. Another important direction is investigating how LLM biases and stochasticity may affect large-scale simulations.
Overall, this work contributes to bridging LLM-based reasoning and crowd simulation, opening new avenues for more believable, realistic, and adaptive virtual human behavior modeling.

%PA: future work ideas:
% - synthetic vision
% - more LLM options
% - more personality variation
% - communication between agents

\section*{Acknowledgements}
%RU: CNPq (Soraia), INCT (Paulo, Rubens e Schneider) e Kunumi (Rubens e Schneider).
%PA: checa se o Stefano tem algo. Acho que não, mas só para ter certeza
This study was partly financed by the Coordenação de Aperfeiçoamento de Pessoal de Nivel Superior – Brazil (CAPES) – Finance Code 001, by the Conselho Nacional de Desenvolvimento Científico e Tecnológico - Brazil (CNPq),  by Conselho Nacional de Desenvolvimento Científico e Tecnológico - Brazil (INCT SiMAI, CNPq \#408330/2024-4), and by Kunumi Institute. The authors thank the institutions for their financial support and commitment to advancing scientific research. We would like to thank Alice Cestari Colares for modeling the virtual office floor used in this work.

% For the bibliography, use the standard 'unsrt' package, provided with LaTeX
% We prefer the usage of '.bib' files, since they are style independent (see
% the bibliography file 'example.bib' for an example of bib definitions)

\bibliographystyle{unsrt}
\bibliography{example}

\end{document}